\documentclass[epj,nopacs]{svjour}

\usepackage{amsmath}
\usepackage{amssymb}
\usepackage{amsfonts}
\usepackage{graphicx}
\usepackage{dcolumn}
\usepackage{bm}
\usepackage{mathrsfs}
\usepackage{accents}
\usepackage[inline]{enumitem}
\usepackage[colorlinks=true, pdfstartview=FitV, linkcolor=blue, citecolor=blue, urlcolor=blue]{hyperref} 

\newcommand{\ac}{\accentset}

\begin{document}

\title{Constraining the attractive fifth force in the general free scalar-tensor gravity with solar system experiments}
\titlerunning{Constraining the attractive fifth force in the general free scalar-tensor gravity ...}

\author{Xing Zhang\inst{1,2,3}\thanks{e-mail: zhxing@nwu.edu.cn}
\and Bo Wang\inst{4,5}
\and Rui Niu\inst{4,5}
}

\institute{School of Physics, Northwest University, Xi’an 710127, China
\and Shaanxi Key Laboratory for Theoretical Physics Frontiers, Xi’an 710127, China
\and Peng Huanwu Center for Fundamental Theory, Xi’an 710127, China
\and CAS Key Laboratory for Researches in Galaxies and Cosmology, Department of Astronomy, University of Science and Technology of China, Chinese Academy of Sciences, Hefei 230026, China
\and School of Astronomy and Space Sciences, University of Science and Technology of China, Hefei 230026, China
}

\date{Received: date / Revised version: date}

\abstract{
In this paper, we focus on the general free scalar-tensor gravity with three free coupling functions, which in the near-field region looks like general relativity (GR) plus a fifth force of Yukawa-type induced by the scalar field.
We show that the fifth force is always attractive in the theory.
We investigate the effects of the attractive fifth force and calculate in detail the fifth force-induced orbital precession rate $\delta\omega/\omega$ and the parameterized post-Newtonian parameters $\gamma$ and $\beta$, all of which depend on the fifth force parameters and the interaction distance.
It turns out that, due to the attractive fifth force, $\delta\omega/\omega$ is always greater than zero, $\gamma$ is always less than one, $\beta$ is greater than one at large distances, and additionally this class of theories is ruled out as an alternative theory to dark matter.
We place stringent constraints on the fifth force parameters by combining the lunar laser ranging (LLR), Cassini, and Mercury precession experiments, and derive the upper bounds on the strength ratio of the fifth force to gravitational force at different scales from the LLR observation.
We find that the Mercury constraint is not competitive with the LLR and Cassini constraints and the LLR observation imposes much more stringent bounds on the strength ratio on large scales than on small scales.
Our results show that this theory is sufficiently close to GR for a small enough fifth force strength and can reduce to GR with a minimally coupled scalar field in the absence of fifth force.
} 

\maketitle

\section{Introduction}\label{sec1_Intro}
Although Einstein's general relativity (GR) has been very successful at interpreting gravity, it has been plagued by the problems of quantization \cite{Kiefer:2007aa,DeWitt:1967yk} as well as dark matter and dark energy  \cite{Cline:2013aa,Sahni:2004ai}.
It is commonly believed by scientists that GR is not the final theory of gravitational interaction. 
Therefore, studies of alternative theories of gravity play a crucial role in testing GR \cite{Capozziello:2011et,Clifton:2012aa}.
It is now known that GR is the unique interacting theory of a Lorentz invariant massless spin-2 particle \cite{Weinberg:1965wb,Weinberg:1980vt}.
Based on this, a natural way to extend GR is to add extra degrees of freedom.
Scalar fields are widely used in the fundamental physics \cite{Becker:2006ug} and in cosmology \cite{Weinberg:1989vw,Guth:1981aa,Bassett:2006uu,Joyce:2015aa}, and additionally the discovery of Higgs boson \cite{ATLAS-Collaboration:2013vu} has shown that scalar particles really exist in nature.
Based on these considerations, scalar degrees of freedom are introduced into most alternative theories of gravity, and the simplest one is the scalar-tensor gravity \cite{Brans:1961aa,Bergmann:1968aa}, which can generally generate a fifth force \cite{Fischbach:1999aa,Adelberger:2003us} to mediate gravitational interactions.

Scalar-tensor theories of gravity have been extensively studied on a huge range of scales, from laboratory tests \cite{Fischbach:1999aa,Adelberger:2003us,Adelberger:2009aa}, to solar system \cite{Damour:1994wi,Capozziello:2010ug,Capozziello:2005vw,Benisty:2022uw,Ursulov:2017tw,Iorio:2007wf,Zhang:2016aa} and binary pulsar \cite{Damour:1993tp,Zhang:2017aa,Zhang:2019ab,Zhang:2022vb} tests, to galactic \cite{Cardone:2011vy,Stabile:2013aa,Cervantes-Cota:2007td,Cervantes-Cota:2009wc} and cosmological \cite{Damour:1993ua,Faraoni:2004pi,Cai:2010te} tests.
However, the literature cited above only studied the scalar-tensor gravity with one or two coupling functions, which is not the most general case.
In this paper, we focus on the general free scalar-tensor gravity, which contains three free coupling functions, namely a free coupling function controlling the strength of the interaction, a free scalar field kinetic coupling function, and a free scalar potential function.

We first consider the simple case of a vanishing scalar potential.
In the case, for a perfect fluid as a matter source, we derive the 2 post-Newtonian (PN), 3PN and 4PN equations for the massless tensor and massless scalar fields, respectively, and calculate in detail all ten parameterized post-Newtonian (PPN) parameters by solving these PN equations.
We find that in the case, the fifth force of Yukawa-type is absent, the PPN parameters $\gamma$ and $\beta$ are model-dependent constants, and the remaining eight PPN parameters are all zero.

In particular, we focus on the theory of a non-vanishing free scalar potential.
In the theory, by solving the 2PN and 4PN equations of the massless tensor and massive scalar fields for a point-like matter source, we derive the effective gravitational constant $G_{\rm eff}$ and the PPN parameters $\gamma$ and $\beta$.
We find that in the theory the fifth force of Yukawa-type appears and the three parameters mentioned above are no longer constants but depend on the interaction distance from the test mass to the gravitational source.
The fifth force affects the orbital dynamics of the system, slightly breaking the inverse-square law of Newtonian gravity and thus inducing the perihelion precession of the orbit.
We derive the perihelion precession induced by any deviation from the inverse-square law by starting from the Lagrangian of the orbital motion and using perturbation analyses.
We obtain the fifth force-induced perihelion precession by reducing any deviation to the Yukawa-type fifth force.
We find that the coupling strength $\alpha$ and mediator mass $m_0$ of the Yukawa-type fifth force can be represented as the certain combinations of three free functions of the theory and can be used to more concisely re-express the PPN parameters $\gamma$ and $\beta$.
By analyzing the functional form of the coupling strength $\alpha$, we find that in the theory the fifth force of Yukawa-type is surprisingly always attractive.
This leads to the following interesting conclusions for the general free scalar-tensor gravity.
First, this theory (with an attractive Yukawa-type fifth force) is insufficient to explain the flat rotation curves of galaxies without the need for dark matter (building on previous works in \cite{Sanders:1984ua,Sanders:1986vc,Sanders:1986tz,Sanders:1990uj}).
Second, the fifth force-induced perihelion precession is always in the same direction as the orbital motion.
Third, the PPN parameter $\gamma$ is always less than one, while $\beta$ is greater than one at large distances, and they both tend to 1 in the limit of large distances.
In short, although not strictly true, this is a good physical picture: in the near-field region, the general free scalar-tensor gravity looks like GR plus an attractive fifth force of Yukawa-type.

Finally, we place stringent constraints on the fifth force parameter space $(\alpha, m_0)$ by combining the experiments of lunar laser ranging (LLR) \cite{Williams:2004aa}, Cassini \cite{Bertotti:2003aa}, and Mercury perihelion precession \cite{Will:2014aa}.
We find that all values of the coupling strength are allowed when the mediator mass is greater than a certain value that is different for different experimental constraints.
The constraint from Mercury precession is the weakest one of all.
As the mediator mass decreases, the Cassini constraint becomes tighter and tighter and tends to a constant.
For small mediator masses the LLR constraint is not competitive with the Cassini constraint, but as the mediator mass increases the LLR constraint becomes tighter and tighter until it reaches the tightest constraint. 
Furthermore, from the LLR observation, we derive the upper bounds on the fifth force to gravitational force strength ratio $f_5(r)$ at different scales.
It is an interesting result that the upper bounds on the strength ratio $f_5(r)$ become tighter and tighter as the interaction distance increases.

The organization of this paper is as follows. 
In Sec. \ref{sec2_gfst}, we display the action and field equations for the general free scalar-tensor gravity.
In Sec. \ref{sec3_10ppn}, we calculate in detail all ten PPN parameters for the theory in the simple case of a vanishing scalar potential.
In Sec. \ref{sec4_2ppn}, we focus on the theory with a non-vanishing free scalar potential and calculate in detail the effective gravitational constant and the PPN parameters $\gamma$ and $\beta$ for the theory.
In Sec. \ref{sec5_f5}, we study the Yukawa-type fifth force in the theory and derive in detail the fifth force-induced perihelion precession.
In Sec. \ref{sec6_prop}, we discuss some interesting properties of the general free scalar-tensor gravity.
In Sec. \ref{sec7_const}, we place stringent constraints on the parameter space of the attractive fifth force by solar system experiments and derive the upper bounds on the scalar to tensor force strength ratio at different distances from the LLR observation.
We conclude in Sec. \ref{sec8_concl}.

Throughout this paper, the metric convention is chosen as $(-,+,+,+)$, Greek indices ($\mu,\nu,\cdots$) run over $0,1,2,3$, and the units $c=\hbar=1$ are adopted.

\section{General free scalar-tensor gravity}\label{sec2_gfst}
A general free scalar-tensor gravity with three free coupling functions is given by the following action \cite{Bergmann:1968aa,Brans:1961aa}:
\begin{eqnarray} \label{ST_V_action}
S &=&
\frac{1}{16\pi G} \int d^4x\sqrt{-g} \big[F(\phi)R 
- W(\phi)(\nabla\phi)^2 
\nonumber\\
&&- V(\phi)\big] + S_m \left(\psi_m, g_{\mu\nu}\right),
\end{eqnarray}
where $G$ is the bare gravitational constant, $g$ is the determinant of the metric $g_{\mu\nu}$, $R$ is the Ricci scalar, and $S_m$ is the action of matter fields $\psi_m$.
$F(\phi)$ is the coupling function that controls the local value of the gravitational coupling constant, $W(\phi)$ is the coupling function that describes the scalar field kinetic interaction, and $V(\phi)$ is the scalar field potential function that endows the scalar field with mass.
Here, the three coupling functions $F(\phi)$, $W(\phi)$ and $V(\phi)$ are all free functions, and they or their combinations are not constrained by theoretical considerations.

The variation of the action \eqref{ST_V_action} with respect to the metric and scalar fields yields the dynamical equations of $g_{\mu\nu} $ and $\phi$:
\begin{eqnarray} \label{Eqs_t1}
&&
F(R_{\mu\nu} \!-\! \frac{1}{2}R g_{\mu\nu}) 
+ \frac{1}{2}Vg_{\mu\nu} 
+ \frac{1}{2}(W \!+\! 2F'')(\partial\phi)^2 g_{\mu\nu}
\nonumber\\
&&
- (W \!+\! F'')\partial_{\mu}\phi\partial_{\nu}\phi 
\!-\! F'(\nabla_{\nu}\partial_{\mu}\phi \!-\! g_{\mu\nu}\square\phi)
\!=\! 8\pi GT_{\mu\nu} ,\,\;
\end{eqnarray}
\begin{eqnarray} \label{Eqs_s1}
(3{F'}^2+2FW)\square\phi 
+((FW)'+3F'F'')(\partial\phi)^2
\nonumber\\
+(2F'V-FV')
=8\pi GF'T ,
\end{eqnarray}
where a prime ($'$) denotes differentiation with respect to $\phi$, $T_{\mu\nu} \equiv (-2/\sqrt{-g})\delta S_m/\delta g^{\mu\nu}$ is the energy-momentum tensor of matter fields, and $T$ is its the trace.

\section{All ten PPN parameters for $V = 0$}\label{sec3_10ppn}
In order to distinguish different theories of gravity and to compare their predictions with a large number of experiments, the PPN formalism has been developed \cite{Eddington:1923ty,Robertson:1962ub,Schiff:1967uk,Nordtvedt:1968ab,Will:1971aa,Will:1993aa,Will:2014aa}.
In the PPN formalism, the PN metrics for gravity theories are parameterized by a set of PPN parameters, which have been measured by various high precision experiments \cite{Will:2014aa}.

In this section, we use the PPN formalism to derive in detail all ten PPN parameters of the general free scalar-tensor gravity with a vanishing scalar potential for a perfect fluid as a matter source.
In the PPN formalism, the gravitational field of the matter source is weak (i.e., $GM/r\ll1$) and its typical velocity $\bm v$ is small (i.e., $v^2\sim GM/r\ll1$), and thus the PN metrics can be calculated by expanding the field equations to $\mathcal{O}(v^n)$ and solving them.

In the weak-field, the tensor and scalar fields can be expanded around a flat Minkowski metric $\eta_{\mu\nu}=diag(-1,1,1,1)$ and a scalar background $\phi_0$ as follows:
\begin{eqnarray} \label{weak_field_st}
g_{\mu\nu} = \eta_{\mu\nu} + h_{\mu\nu} ,\;\;\;
\phi = \phi_0+\varphi \,,
\end{eqnarray}
where the perturbations $h_{\mu\nu}$ and $\varphi$ can be expanded to $\mathcal{O}(v^n)$ and are denoted by superscript $(n)$ below.

For a perfect fluid as a matter source, its energy-momentum tensor is given by \cite{Will:1993aa,Will:2014aa}
\begin{eqnarray}\label{Tuv_per_fluid}
T^{\mu\nu} = \left( \rho + \rho \Pi + p \right) u^\mu u^\nu + p g^{\mu\nu},
\end{eqnarray}
where $u^\mu$ is the fluid four-velocity, $\rho$ is the rest mass density of the fluid, $p$ is the fluid pressure, and $\Pi$ is the internal energy per unit rest mass.

In order to solve the field equations in the PPN formalism, we use the PN gauge introduced in \cite{Will:1993aa,Will:2014aa},
\begin{eqnarray} \label{PN_gauge0123}
h^{\mu}_{\nu, \mu} - \frac{1}{2} h^{\mu}_{\mu, \nu} - \frac{1}{2} \eta_{0\nu} h_{00,0}
= ( F'F^{-1} )_0 \varphi_{, \nu} \,.
\end{eqnarray}
By using Eqs.~\eqref{weak_field_st}, \eqref{Tuv_per_fluid} and \eqref{PN_gauge0123} and expanding the dynamical equations \eqref{Eqs_t1} and \eqref{Eqs_s1} (set $V=0$) to $\mathcal{O}(v^2)$, we can obtain the 2PN equations of the fields $\varphi$, $h_{00}$ and $h_{ij}$ in the form
\begin{eqnarray}
\nabla^2 \ac{(2)}{\varphi} 
&=&  - 8 \pi \rho  G  \Big( \frac{ F' }{ 3{F'}^2 + 2FW } \Big)_0 ,
\\
\nabla^2 \ac{(2)}h_{00} 
&=&  - 16 \pi \rho \frac{ G }{F_0} \Big( \frac{ 2{F'}^2 + FW }{ 3{F'}^2 + 2FW } \Big)_{0} , 
\\
\nabla^2 \ac{(2)}h_{ij} 
&=&  - 16 \pi  \rho \delta_{ij}  \frac{ G }{F_0}  \Big( \frac{ {F'}^2 + FW }{ 3{F'}^2 + 2FW } \Big)_{0} ,
\end{eqnarray}
where the subscript `0' indicates the value of a quantity at $\phi_0$.
These are Poisson's equations and the solutions are
\begin{eqnarray}
\label{varphi_2}
\ac{(2)}\varphi  &=&  \Big( \frac{ F'F }{ 2{F'}^2 + FW } \Big)_0 G_{\rm N} U,        \\
\label{h_00_2}
\ac{(2)}h_{00}   &=&  2 G_{\rm N} U,           \\
\label{h_ij_2}
\ac{(2)}h_{ij}     &=&  2  \Big( 1 - \frac{ {F'}^2 }{ 2{F'}^2 + FW } \Big)_0  G_{\rm N} U \delta_{ij},  
\end{eqnarray}
where $U$ is the Newtonian potential (see Eqs.~\eqref{Metric_potentials}). 
Here, the quantity 
\begin{eqnarray} \label{G_N}
G_{\rm N}
= \frac{G (4{F'}^2+2FW)}{F(3{F'}^2+2FW)} \Big|_{\phi_0}
\end{eqnarray}
can be interpreted as the Newtonian gravitational constant and is set to one in the subsequent calculations.
From the above equation, we see that the fifth force mediated by the scalar field is absent in the general free scalar-tensor gravity with a vanishing scalar potential.

In the PN gauge \eqref{PN_gauge0123}, expanding up to $\mathcal{O}(v^3)$, the time-space component of the metric field equation \eqref{Eqs_t1} is expressed in the form
\begin{eqnarray} \label{h_0j_eq}
\nabla^2\ac{(3)}h_{0j} + \frac{1}{2}\ac{(2)}h_{00,0j} 
= 8\pi \Big( 2 - \frac{ {F'}^2 }{ 2{F'}^2 + FW } \Big)_0 \rho v_j ,
\end{eqnarray}
using Eqs.~\eqref{h_00_2} and \eqref{Rel_7} the solution is 
\begin{eqnarray} \label{h_0j}
\ac{(3)}h_{0j}
= - \frac{1}{2} \Big( 7 - \frac{ 4{F'}^2 }{ 2{F'}^2 + FW } \Big)_0 V_j - \frac{1}{2}W_j ,
\end{eqnarray}
where $V_j$ and $W_j$ are the metric potentials, defined in Eqs.~\eqref{Metric_potentials}.

Using the PN gauge in Eq.~\eqref{PN_gauge0123} and $T^{\mu\nu}$ in Eq.~\eqref{Tuv_per_fluid}, and expanding the time component of the metric field equation \eqref{Eqs_t1} to $\mathcal{O}(v^4)$, we obtain the 4PN equations of the field $h_{00}$,
\begin{eqnarray} \label{h_00_4_eq}
&&   \nabla^2\ac{(4)}h_{00} + (\nabla\ac{(2)}h_{00})^2 
                 - \ac{(2)}h_{jk}\ac{(2)}h_{00,jk}                        \nonumber\\
&&   + \frac{( F'(FW)' - 2 F''FW )_0}{( 3 {F'}^2 F + 2F^2W )_0} (\nabla\ac{(2)}\varphi)^2    \nonumber\\
&=& - 4\pi \Big\{ 2\rho \Pi 
                 + 2\Big(2 - \frac{{F'}^2}{2{F'}^2+FW}\Big)_0 \rho v^2       \nonumber\\
&&   + 6 \Big(1 - \frac{{F'}^2}{2{F'}^2+FW}\Big)_0 p                        \nonumber\\
&&   + \Big[\frac{ 2F{F'}^2( 2F''FW - F'(FW)' ) }{ (2{F'}^2 + FW)^2 (3{F'}^2 + 2FW) }  \nonumber\\
&&   - \frac{ 2{F'}^2 }{ 2{F'}^2 + FW } - 4 \Big]_0 \rho U \Big\} \,.
\end{eqnarray}
Substituting Eqs.~\eqref{varphi_2}, \eqref{h_00_2} and \eqref{h_ij_2} into the above equation and using the relations \eqref{Rel_8} and \eqref{Rel_9}, we obtain the solution in the form 
\begin{eqnarray} \label{h_00_4}
\ac{(4)}h_{00}
&=& - 2 \Big[ 1 + \frac{F{F'}^2(F'(FW)'-2F''FW)}{4(2{F'}^2+FW)^2(3{F'}^2+2FW)} \Big]_0 U^2     \nonumber\\
&&   + 2\Big(2 - \frac{{F'}^2}{2{F'}^2+FW}\Big)_0\Phi_1 + 2 \Big[ 2 - \frac{ 3{F'}^2 }{ 2{F'}^2 + FW }     \nonumber\\
&&   - \frac{ F{F'}^2(F'(FW)'-2F''FW) }{ 2(2{F'}^2+FW)^2(3{F'}^2+2FW) } \Big]_0 \Phi_2 + 2\Phi_3     \nonumber\\
&&   + 6 \Big(1 - \frac{{F'}^2}{2{F'}^2+FW}\Big)_0 \Phi_4 \,,
\end{eqnarray}
where $\Phi_1$, $\Phi_2$, $\Phi_3$ and $\Phi_4$ are the metric potentials, defined in Eqs.~\eqref{Metric_potentials}.

By comparing the metric perturbations in Eqs.~\eqref{h_00_2}, \eqref{h_ij_2}, \eqref{h_0j} and \eqref{h_00_4} with the standard form of the PPN metric in Eqs.~\eqref{PPN_metric}, we can identify all ten PPN parameters in the form
\begin{eqnarray}
\label{ST_PPNg}
\gamma  &=&  1 - \frac{{F'}^2}{2{F'}^2+FW}\Big|_{\phi_0} ,       \\
\label{ST_PPNb}
\beta      &=&  1 + \frac{ F{F'}^2(F'(FW)'-2F''FW) }{ 4(2{F'}^2+FW)^2(3{F'}^2+2FW) } \Big|_{\phi_0} ,       \\
\label{ST_8PPN}
\xi          &=&  \alpha_1 = \alpha_2 = \alpha_3 = \zeta_1 = \zeta_2 = \zeta_3 = \zeta_4 = 0 ,    
\end{eqnarray}
where the significances of these parameters are given in Appendix \ref{app1_ppn}.

As we can see, in the case of a vanishing scalar potential, the PPN parameters $\gamma$ and $\beta$ are model-dependent constants, and the remaining PPN parameters are all zero.
However, in the theory with a non-vanishing free scalar potential, due to the emergence of the Yukawa-type fifth force, these two parameters $\gamma$ and $\beta$ are no longer constants but distance-dependent, as we will discuss in the next section.

\section{Effective gravitational constant and PPN parameters $\gamma$ and $\beta$ for $V \neq 0$}\label{sec4_2ppn}
Now let us focus on the general free scalar-tensor gravity with three free functions in the case of a point mass $M$ as a matter source, which can be described by the energy-momentum tensor~\eqref{Tuv_per_fluid} with $\rho = M\delta(r)$ and vanishing $v_i$, $p$ and $\Pi$ .

\subsection{Effective gravitational constant $G_{\rm eff}$ and PPN parameter $\gamma$}
In the PPN formalism, imposing the PN gauge \eqref{PN_gauge0123} and expanding the dynamical equations \eqref{Eqs_t1} and \eqref{Eqs_s1} to $\mathcal{O}(v^2)$, we obtain the 2PN equations of the fields $\varphi$, $h_{00}$ and $h_{ij}$ in the form
\begin{eqnarray} 
\label{Eqs_s2}
&&( \nabla^2 - m_s^2 ) \ac{(2)}{\varphi} = - 8 \pi \rho  G  \Big( \frac{ F' }{ 3{F'}^2 + 2FW } \Big)_0 \,,
\\
\label{Eqs_t2}
&&\nabla^2 \ac{(2)}h_{00} - \Big( \frac{F'}{F} \Big)_0 m_s^2 \ac{(2)}{\varphi} 
= - 16 \pi \rho \frac{ G }{F_0} \Big( \frac{ 2{F'}^2 \!+\! FW }{ 3{F'}^2 \!+\! 2FW } \Big)_{\!0} ,  \quad
\\
\label{Eqs_t3}
&&\nabla^2 \ac{(2)}h_{ij} \!+\! \Big(\! \frac{F'}{F} \!\Big)_{\!0} \!  m_s^2 \ac{(2)}{\varphi} \delta_{ij}
\!=\! - 16 \pi \! \rho \delta_{ij} \! \frac{ G }{F_0} \! \Big(\! \frac{ {F'}^2 \!+\! FW }{ 3{F'}^2 \!+\! 2FW } \!\Big)_{\!0} , \quad
\end{eqnarray}
where the quantity $m_s$ can be interpreted as the effective mass of the scalar field, and given by
\begin{eqnarray} \label{scalar_mass_eff}
m_s \equiv
\Big( \frac{ F V'' }{ 3{F'}^2 + 2FW } \Big)_0^{\frac{1}{2}} .
\end{eqnarray}
Here, the first is a screened Poisson equation, and the last two can be simplified as the Poisson equations by using the first.
By setting $\rho=M \delta(r)$ and the solutions of these equations are given by
\begin{eqnarray} 
\label{phi2}
\ac{(2)}{\varphi}
&=& \Big( \frac{ 2F' }{ 3{F'}^2 + 2FW } \Big)_0  \frac{ GM }{ r } e^{-m_s r} ,
\\
\label{h002}
\ac{(2)}h_{00}
&=& \frac{2GM}{F_0r} \Big( 1+\frac{{F'_0}^2}{(3{F'}^2+2FW)_0}e^{-m_s r} \Big) ,
\\
\label{hij2}
\ac{(2)}h_{ij}   
&=& \frac{2GM}{F_0r} \delta_{ij} \Big( 1 - \frac{{F'_0}^2}{( 3{F'}^2 + 2FW )_0}e^{-m_s r} \Big) .
\end{eqnarray}

The effective gravitational constant $G_{\rm eff}$ and the PPN parameter $\gamma$ are defined in the form \cite{Will:2014aa},
\begin{eqnarray}
\ac{(2)}h_{00} = \frac{2G_{\rm eff} M}{r},
\qquad
\ac{(2)}h_{ij} = \gamma \frac{2G_{\rm eff} M}{r} \delta_{ij}.
\end{eqnarray}
Using these, from Eqs.~\eqref{h002} and \eqref{hij2}, we obtain 
\begin{eqnarray} \label{G_eff}
G_{\rm eff}
= \frac{G}{F_0} \Big( 1 + \frac{ {F'_0}^2 }{ (3{F'}^2+2FW)_0 }e^{-m_s r} \Big) ,
\end{eqnarray}
\begin{eqnarray} \label{PPN_gamma}
\gamma
= 1 - \frac{ 2 {F'_0}^2 e^{-m_s r} }{ (3{F'}^2+2FW)_0 + {F'_0}^2 e^{-m_s r} } ,
\end{eqnarray}
where $G_{\rm eff}$ and $\gamma$ both are distance-dependent and can reduce to $G/F_0$ and $1$ in the limit of $m_s r \to \infty$ and to Eqs.~\eqref{G_N} and \eqref{ST_PPNg} in the limit of $m_s r \to 0$.
It is clear that the Yukawa exponential factor has been restored since the scalar field is massive.
This means that the Yukawa-type fifth force appears in the theory, as we will see in more detail in the next section.

\subsection{PPN parameter $\beta$}
By imposing the PN gauge \eqref{PN_gauge0123} and the scalar field equation \eqref{Eqs_s1} is expanded up to $\mathcal{O}(v^4)$ in the form
\begin{eqnarray} \label{phi4_eq}
&&    ( \nabla^2 - m_s^2 ) \ac{(4)}{\varphi} =                \nonumber\\
&&    \Big(\! \frac{V'''}{2V''} \!-\! \frac{A'}{A} \!\Big)_{\!0} m_s^2 \ac{(2)}{\varphi}^{\,2} + \Big(\! \frac{F'}{F} - \frac{A'}{2A} \!\Big)_{\!0} ( \nabla \ac{(2)}{\varphi} )^2 + \ac{(2)}{h}_{ij} \ac{(2)}{\varphi}_{,ij}           \nonumber\\
&&    + \ac{(2)}{\varphi}_{,00}  + 8\pi G \rho \Big[ \Big(\! \frac{F'}{A} \!\Big)_{\!0} \! \Big(\! \frac{3p}{\rho} \! - \! \Pi \!\Big) + \Big(\! \frac{A' F'}{A^2} \! - \! \frac{F''}{A} \!\Big)_{\!0} \ac{(2)}{\varphi} \Big] \,,\qquad
\end{eqnarray}
where $A = 3{F'}^2 + 2FW $.
The term $\ac{(2)}{\varphi}_{,00}$ can be dropped for static solutions.
The pressure $p$ and specific internal energy $\Pi$ can be dropped for a point-like matter source.  
The term $\rho\ac{(2)}{\varphi}$ corresponds to gravitational self-energies and can be neglected.
Using $\ac{(2)}{\varphi}$ and $\ac{(2)}{h}_{ij}$ in Eqs.~\eqref{phi2} and \eqref{hij2}, the solution of the remaining equation is given by 
\begin{eqnarray} \label{phi4}
\ac{(4)}{\varphi} = 
&&     \Big( \frac{F'}{A} \Big)^2_0 \Big( \frac{2F'}{F} - \frac{A'}{A} \Big)_0 \Big( \frac{GM}{r} e^{-m_s r} \Big)^2       \nonumber\\
&&     - 3 \Big( \frac{F'}{A} \Big)^2_0 \Big( \frac{A'}{2 A} \!+\! \frac{F'}{F} \!-\! \frac{V'''}{3V''} \Big)_0 m_s r \Big( \frac{GM}{r} e^{-m_s r} \Big)^2       \nonumber\\
&&     \times [ e^{3m_s r} {\rm Ei}(\! - 3m_s r \!)  \!-\! e^{\! m_s r} {\rm Ei}(\! - m_s r \!) ]  + \! 2 \Big(\! \frac{F'}{A F} \!\Big)_{\!0} \! m_s r      \nonumber\\
&&     \times \Big(\! \frac{GM}{r} e^{\!-m_s r} \!\Big)^{\!\!2} [ e^{3m_s r} {\rm Ei}(\!-2m_s r\!) \!-\! e^{m_s r}\! \ln(m_s r) ],   \qquad
\end{eqnarray}
where ${\rm Ei}(x)$ is the exponential integral function, defined by
\begin{align}\label{Ei_x}
\text{Ei}(x) \equiv - \int_{-x}^{\infty} \frac{e^{-t}}{t} dt\,.
\end{align}

In the PN gauge \eqref{PN_gauge0123}, expanding the metric field equation \eqref{Eqs_t1} to $\mathcal{O}(v^4)$, we obtain the 4PN equation of the field $h_{00}$ in the form
\begin{eqnarray} \label{h004_eq}
&&    \nabla^2\ac{(4)}h_{00} + (\nabla\ac{(2)}h_{00})^2 
                   + \Big( \frac{A'F'}{2AF} - \frac{F''}{F} \Big)_0 (\nabla\ac{(2)}\varphi)^2              \nonumber\\
&&    - \ac{(2)}h_{jk}\ac{(2)}h_{00,jk}  + \Big( \frac{F'}{F} \Big)_0 m_s^2 \ac{(2)}h_{00} \ac{(2)}\varphi  
                   - \Big( \frac{F'}{F} \Big)_0 m_s^2 \ac{(4)}\varphi                                             \nonumber\\
&&    - \Big( \frac{ F' V''' }{ 2 F V'' } +  \frac{A}{F} \Big( \frac{F'}{A} \Big)' 
                   + \frac{A}{2F^2} - \frac{{F'}^2}{F^2} \Big)_0 m_s^2 \ac{(2)}\varphi^{\,2}             \nonumber\\
&=&  - \frac{8\pi G}{F_0} \Big[ 2\rho v^2 + \Big( 1 \!+\! \frac{{F'}^2}{A} \Big )_0 \rho \Pi  
                   + 3\Big( 1 \!-\! \frac{{F'}^2}{A} \Big)_0 p                                                              \nonumber\\
&&    - \Big( 1 \!+\! \frac{{F'}^2}{A} \Big )_0 \rho \ac{(2)}h_{00} + \Big( F \Big(\frac{{F'}^2}{AF} \Big)'  
                   \!-\! \frac{F'}{F} \Big)_0 \rho \ac{(2)}\varphi \Big] \,.
\end{eqnarray}
Here, the velocity $v$, pressure $p$ and specific internal energy $\Pi$ can be dropped, since the calculation is performed in the rest frame of the point-like matter source.
The terms $\rho\ac{(2)}{h}_{00}$ and $\rho\ac{(2)}{\varphi}$ can be neglected, since they correspond to gravitational self-energies and do not affect the calculation of the PPN parameter $\beta$.
Using $\ac{(2)}{\varphi}$, $\ac{(2)}{h}_{00}$, $\ac{(2)}{h}_{ij}$ and $\ac{(4)}{\varphi}$ in Eqs.~\eqref{phi2}, \eqref{h002}, \eqref{hij2} and \eqref{phi4}, the remaining equation is solved by 
\begin{eqnarray} \label{h004}
\ac{(4)}h_{00} =
&&     - \frac{2}{F^2_0} \Big( 1 + \frac{{F'_0}^2}{A_0} e^{-m_sr} \Big)^2 \Big( \frac{GM}{r} \Big)^2
\nonumber\\
&&     - \frac{ {F'}^2 (A'F' - 2AF'') }{ A^3 F } \!\Big|_0 \Big( \frac{GM}{r} e^{-m_sr} \Big)^2
\nonumber\\
&&     + \Big(\! \frac{ {F'}^2 }{ AF^2 } \!\Big)_{\!0} \! \Big(\! \frac{GM}{r} e^{-m_sr} \!\Big)^{\!2} \! m_sr \Big[ 1 - 2e^{m_sr}\ln(m_sr) 
\nonumber\\
&&     + 2e^{2m_sr} \!( e^{m_sr} \!\!+\! m_sr ) {\rm Ei}(\!-2m_sr\!) \!-\! 3 \Big(\! \frac{FF'}{A} \!\Big)_{\!0} \!\Big(\! \frac{A'}{2A} 
\nonumber\\
&&     + \frac{F'}{F} \!-\! \frac{V'''}{3V''} \!\Big)_{\!0} \! ( e^{3m_{\!s} \! r}{\rm Ei}(\! -3m_{\!s} r \!) \!-\! e^{m_{\!s} \! r}{\rm Ei}(\!-m_{\!s} r \!) ) \Big]. \qquad
\end{eqnarray}
Using the relationship $\ac{(4)}h_{00} = - 2 \beta ( {G_{\rm eff}M}/{r} )^2$ and $G_{\rm eff}$ in Eq.~\eqref{G_eff}, the PPN parameter $\beta$ is given by
\begin{eqnarray} \label{PPN_beta}
\beta = 
&&      1 + \frac{ F{F'}^2(A'F' - 2AF'') }{ 2A( A e^{m_sr} + {F'}^2 )^2 } \Big|_0             \nonumber\\
&&      - \frac{ A{F'}^2 }{ 2( A e^{m_sr} + {F'}^2 )^2 } \Big|_0  m_sr \Big[ 1 - 2e^{m_sr}\ln(m_sr)           \nonumber\\
&&      + 2e^{2m_sr} \!( e^{m_sr} \!\!+\! m_sr ) {\rm Ei}(\!-2m_sr\!) \!-\! 3 \Big(\! \frac{FF'}{A} \!\Big)_{\!0} \!\Big(\! \frac{A'}{2A}         \nonumber\\
&&      + \frac{F'}{F} \!-\! \frac{V'''}{3V''} \!\Big)_{\!0} \! ( e^{3m_sr}{\rm Ei}(\!-3m_sr\!) \!-\! e^{m_sr}{\rm Ei}(\!-m_sr\!) ) \Big]. \qquad
\end{eqnarray}
where this parameter is also distance-dependent and can reduce to 1 and Eq.~\eqref{ST_PPNb} in the cases of $m_s r \to \infty$ and $m_s r \to 0$, respectively.

We will see in the next two sections that these parameters $G_{\rm eff}$, $\gamma$ and $\beta$ in Eqs.~\eqref{G_eff}, \eqref{PPN_gamma} and \eqref{PPN_beta} can be expressed more concisely as functions of the fifth force parameters.

\section{Yukawa-type fifth force and fifth force-induced perihelion precession}\label{sec5_f5}
In this section we study the Yukawa-type fifth force and the fifth force-induced perihelion precession in the general free scalar-tensor gravity.

\subsection{Yukawa-type fifth force}
In order to more clearly describe the fifth force in the theory, by using the combinations of three functions $F$, $W$ and $V$, we define the following two quantities: 
\begin{eqnarray} \label{alpha_cp}
\alpha \equiv \Big( \frac{ {F'}^2 }{ 3{F'}^2 + 2FW } \Big)_0\,,
\end{eqnarray}
\begin{eqnarray} \label{m0_param}
m_0 \equiv \Big( \frac{ F V'' }{ {F'}^2 } \Big)^{\frac{1}{2}}_0\,.
\end{eqnarray}
As we will see later, these two quantities are the coupling strength and mediator mass of the Yukawa-type fifth force.
Here, $F_0 > 0$ is a consequence of the fact that gravity is attractive (see Eq.~\eqref{G_eff} or \eqref{F_N_5}), and $V''_0 \ge 0$ is the non-negative mass squared of the scalar field.
Additionally, the effective mass in Eq.~\eqref{scalar_mass_eff} is non-negative, i.e., $ F_0 V''_0 / ( 3{F'}^2 + 2FW )_0 \ge 0$.
These above relations imply
\begin{eqnarray} \label{alpha_m0}
\alpha \ge 0\,,
\quad
m_0 \ge 0\,.
\end{eqnarray}

Making use of the definitions of $\alpha$ and $m_0$ in Eqs.~\eqref{alpha_cp} and \eqref{m0_param}, the effective mass in Eq.~\eqref{scalar_mass_eff} is expressed as 
\begin{eqnarray} \label{ms_m0}
m_s=\alpha^{\frac{1}{2}} m_0\,,
\end{eqnarray}
and the effective gravitational constant in Eq.~\eqref{G_eff} is rewritten as
\begin{eqnarray} \label{Geff_a}
G_{\rm eff} = \frac{G}{F_0} ( 1 + \alpha e^{-m_s r} )\,.
\end{eqnarray}
Using this, the potential energy describing the gravitational interaction between two point-like objects with masses $m$ and $M$ is given by
\begin{eqnarray} \label{poten_energy}
V(r) =  -\frac{G}{F_0} \frac{Mm}{r} ( 1 + \alpha e^{-m_s r} )\,.
\end{eqnarray}
Differentiating this above, the force is
\begin{eqnarray} \label{F_N_5}
\bm{F}(r) = - \frac{G}{F_0} \frac{Mm}{r^2} \hat{\bm{r}} [ 1 + \alpha (1+m_s r) e^{-m_s r} ]\,,
\end{eqnarray}
where the first term is the gravitational force (tensor force) and the second term is the Yukawa-type fifth force (scalar force).

\subsection{Fifth force-induced perihelion precession}
Because of the presence of the Yukawa-type fifth force, the net force experienced by a planet does not vary exactly as inverse-square, which induces the perihelion precession of the planet orbit in which the major axis of the planet orbit slowly rotates in the orbital plane.

In the weak-field limit, the Lagrangian describing the orbital dynamics of two point-like objects with masses $m$ and $M$ ($m<<M$) in the polar coordinates ($r,\theta$) is given by
\begin{eqnarray} 
L = \frac{1}{2}m(\dot{r}^2 + r^2\dot{\theta}^2) - V(r) ,
\end{eqnarray}
where an overdot denotes differentiation with respect to time.
By varying the Lagrangian and performing the replacement $r=u^{-1}$, the equation of motion is
\begin{eqnarray} \label{eq_motion_orbit_F}
\frac{d^2u}{d\theta^2} + u = - \frac{1}{ml^2u^2} F(u^{-1}) ,
\end{eqnarray}
where $l=r^2\dot{\theta}$ is the (conserved) orbital angular momentum per unit mass.
The force $F(r)\equiv-dV(r)/dr$ can usually be rewritten in the following form
\begin{eqnarray} \label{FN_1f}
F(r) = - \frac{G_{\rm N}Mm}{r^2} [ 1 + f(r) ],
\end{eqnarray}
where $f(r)$ is the sum of various deviations from the inverse-square law of Newtonian gravity.
Substituting the above equation into Eq.~\eqref{eq_motion_orbit_F} yields
\begin{eqnarray} \label{eq_motion_orbit_f}
\frac{d^2u}{d\theta^2} + u = \frac{1}{p}[ 1 + f(u^{-1}) ],
\end{eqnarray}
where $p=l^2/(G_{\rm N}M)$ is the semi-latus rectum of the unperturbed orbit and is linked to the the semi-major axis $a$ and eccentricity $e$ of the unperturbed orbit by $p=a(1-e^2)$.

In the above equation, $f(u^{-1})$ can be regarded as a small perturbation, then the solution should lie near the unperturbed value.
For this reason, expanding the right hand side of Eq.~\eqref{eq_motion_orbit_f} in Taylor's series at $p^{-1}$, we obtain the solution in the form 
\begin{align} \label{u_theta}
u(\theta) = u_p + u_e \cos{\omega (\theta-\theta_0)},
\end{align}
with
\begin{eqnarray} \label{omega_f}
\omega = \Big[1 + p \frac{df(r)}{dr}\Big|_{r=p}\Big]^{\frac{1}{2}} ,
\end{eqnarray}
\begin{eqnarray} 
u_p = \frac{1}{p} \Big[1 + \frac{f(p)}{\omega^2}\Big] ,
\end{eqnarray}
where $u_e=eu_p$ for a quasi-elliptical orbit.
From Eq.~\eqref{u_theta} we see that the perihelion occurs at $\omega (\theta-\theta_0) \equiv \omega\theta_n = 2\pi n $ , $n\in \mathbb{Z}$.
Using this, the precession angle per orbit, $\delta\theta \equiv (\theta_{n+1} - \theta_{n}) - 2\pi$ is given by
\begin{eqnarray} 
\frac{\delta\theta}{2\pi} = \frac{1 - \omega}{\omega} \equiv \frac{\delta\omega}{\omega} .
\end{eqnarray}
Substituting Eq.~\eqref{omega_f} into the above equation and keeping only the leading order, we obtain the perihelion precession rate 
\begin{eqnarray} \label{precession_rate}
\frac{\delta\theta}{2\pi} = \frac{\delta\omega}{\omega} = - \frac{p}{2} \frac{df(r)}{dr}\Big|_{r=p}\,.
\end{eqnarray}
Note that the above relationship is also obeyed between each component of $f(r)$ and its induced precession.

By comparing Eq.~\eqref{F_N_5} and Eq.~\eqref{FN_1f}, we obtain the strength ratio of the fifth force to gravitational force, 
\begin{eqnarray} \label{f5_r}
f_5(r) = \alpha (1+m_s r) e^{-m_s r}\,,
\end{eqnarray}
and this is also the strength ratio of the scalar to tensor force in the general free scalar-tensor gravity.
Substituting this into Eq.~\eqref{precession_rate}, the fifth force-induced perihelion precession rate is given by
\begin{eqnarray} \label{f5_precession}
\frac{\delta\theta_5}{2\pi} = \frac{\delta\omega_5}{\omega} = \frac{\alpha}{2} (m_s p)^2 e^{- m_s p} \,.
\end{eqnarray}
Here, we note that in the presence of the Yukawa-type fifth force, as $m_0\to 0$ (see Eq.~\eqref{ms_m0}), the induced precession goes to zero.

\section{Properties of general free scalar-tensor gravity}\label{sec6_prop}
In this section we summarize and discuss some interesting properties of the general free scalar-tensor gravity.

Making use of the fifth force parameters $\alpha$ and $m_0$ in Eqs.~\eqref{alpha_cp} and \eqref{m0_param}, the PPN parameter $\gamma$ in Eq.~\eqref{PPN_gamma} can be rewritten as
\begin{eqnarray} \label{gamma_a}
\gamma
= \frac{ 1 - \alpha e^{-m_s r} }{ 1 + \alpha e^{-m_s r} }\,,
\end{eqnarray}
and when $m_s r > 1$ the PPN parameter $\beta$ in Eq.~\eqref{PPN_beta} can be approximated as
\begin{eqnarray} \label{beta_a}
\beta
= 1 + \frac{ \alpha e^{m_s r} [ 1 + 2 m_s r \ln(m_s r) ] }{ 2(\alpha + e^{m_s r})^2 }\,.
\end{eqnarray}
We now get the four important Eqs.~\eqref{f5_r}, \eqref{f5_precession}, \eqref{gamma_a} and \eqref{beta_a} in this paper.
Our main results are also summarized in these equations.
According to Eq.~\eqref{alpha_m0}, the coupling strength $\alpha$ is always non-negative.
Now let us consider the two cases of $\alpha=0$ and $\alpha>0$.

In the case of $\alpha=0$ (i.e., $F'=0$), the coupling function $F$ is a constant, the non-minimal coupling disappears, and we have the following results:
(i)~$f_5(r)=0$ and $\delta\omega_5/\omega=0$, i.e., the Yukawa-type fifth force and its induced precession are absent;
(ii)~the PPN parameters $\gamma = \beta = 1$.
These indicate that the theory in the case reduces to GR with a minimally coupled scalar field.

In the case of $\alpha > 0$ (i.e., $F'\neq0$), the presence of the non-minimal coupling makes the theory non-trivial, and we find the following interesting results for the general free scalar-tensor gravity:
\begin{enumerate}[label = (\roman*)]
\item\label{item_1}   $f_5(r)>0$, i.e., the Yukawa-type fifth force in the theory is always attractive;
\item\label{item_2}  $\delta\omega_5/\omega>0$, i.e., the precession is in the same direction as the orbital motion;
\item\label{item_3}  $\gamma < 1$, and $\beta > 1$ for $m_s r > 1$.
\end{enumerate}
Note that, according to the previous works
\footnote{Refs.~\cite{Sanders:1984ua,Sanders:1986vc,Sanders:1986tz,Sanders:1990uj} have shown that an almost constant profile of galactic rotation curve is recovered by adding a repulsive Yukawa-type fifth force to gravitational force.},
the result \ref{item_1} means that the general free scalar-tensor gravity is not sufficient to reproduce the flat rotation curves of galaxies without the need for dark matter, that is, the theory cannot be used as an alternative theory to dark matter.
The results \ref{item_2} and \ref{item_3} show that the experimental constraints on the theory come only from the upper bound on $\delta\omega/\omega$, the lower bound on $\gamma$ and the upper bound on $\beta$.

\section{Solar system constraints}\label{sec7_const}
In this section we place constraints on the parameter space of the attractive fifth force in the theory by solar system experiments involving the LLR measurement \cite{Williams:2004aa}, the Cassini satellite \cite{Bertotti:2003aa}, and the perihelion precession of Mercury \cite{Will:2014aa}.

The perihelion precession is one of the four classic solar system tests of GR.
Bounds on anomalous precession deviating from GR in the solar system provide extremely stringent constraints on the fifth force, in which the most stringent bound comes from the LLR observation \cite{Williams:2004aa} searching for the anomalous precession of the Moon.
The Moon moves on an elliptical orbit with a small eccentricity of $0.0549$ and a semi-major axis of $0.00257$ astronomical unit ($\rm AU\simeq 1.496\times 10^8 km$).
In the Earth-Moon system, the characteristic interaction distance is the semi-latus rectum of the orbit, i.e., $0.00256 \rm AU$.
The observed quantities relevant to placing constraints are presented in Table \ref{tab_1}.

The PPN parameters have been measured accurately by a large number of experiments \cite{Will:2014aa}. 
The current most stringent constraints on the PPN parameters $\gamma$ and $\beta$ come from the Cassini measurements of the Shapiro time-delay \cite{Bertotti:2003aa} and the perihelion precession of Mercury \cite{Will:2014aa}, respectively.
These two experiments also provide a characteristic interaction distance, respectively.
In the Cassini measurements \cite{Bertotti:2003aa}, the radio signals sent between the Cassini spacecraft and the Earth were passing by the Sun at a distance of $0.00744 \rm AU$, which is the characteristic interaction distance in the experiment.
In the perihelion precession of Mercury \cite{Will:2014aa}, Mercury moves on an elliptical orbit with a moderate eccentricity of $0.2056$ and a semi-major axis of $0.3871 \rm AU$, and the characteristic interaction distance is the semi-latus rectum, i.e., $0.3707\rm AU$.
The measured values of the relevant parameters are presented in Table \ref{tab_1}.

\begin{table}[!htbp]
\centering
\caption{Current experimental constraints on the Moon perihelion precession rate ${\delta\omega/\omega}$ and the PPN parameters $\gamma$ and $\beta$.}
\label{tab_1}
\resizebox{0.48\textwidth}{1.2cm}{
\begin{tabular}{p{1.1cm} < {\raggedright} p{2.7cm} < {\centering} p{2.4cm}<{\centering} p{1.8cm}<{\raggedleft}}
\hline\hline
Param- eters                    &  Measured value                             &  Interaction distance, $d_i$ (\!AU\!)  &  Experiments  \\
\hline
${\delta\omega/\omega}$ &  $(-2.1 \pm 7.1) \!\times\! 10^{-12}$ &  0.00256                                      &   LLR \cite{Williams:2004aa} \\
$\gamma-1$                    &  $(2.1 \pm 2.3) \!\times\! 10^{-5}$   &  0.00744                                      &   Cassini \cite{Bertotti:2003aa} \\
$\beta-1$                        &  $(-4.1 \pm 7.8) \!\times\! 10^{-5}$  &  0.3707                                        &  Mercury \cite{Will:2014aa} \\
\hline \hline
\end{tabular}
}
\end{table}

\begin{figure}[!htbp]
\centering
\includegraphics[width=1\columnwidth]{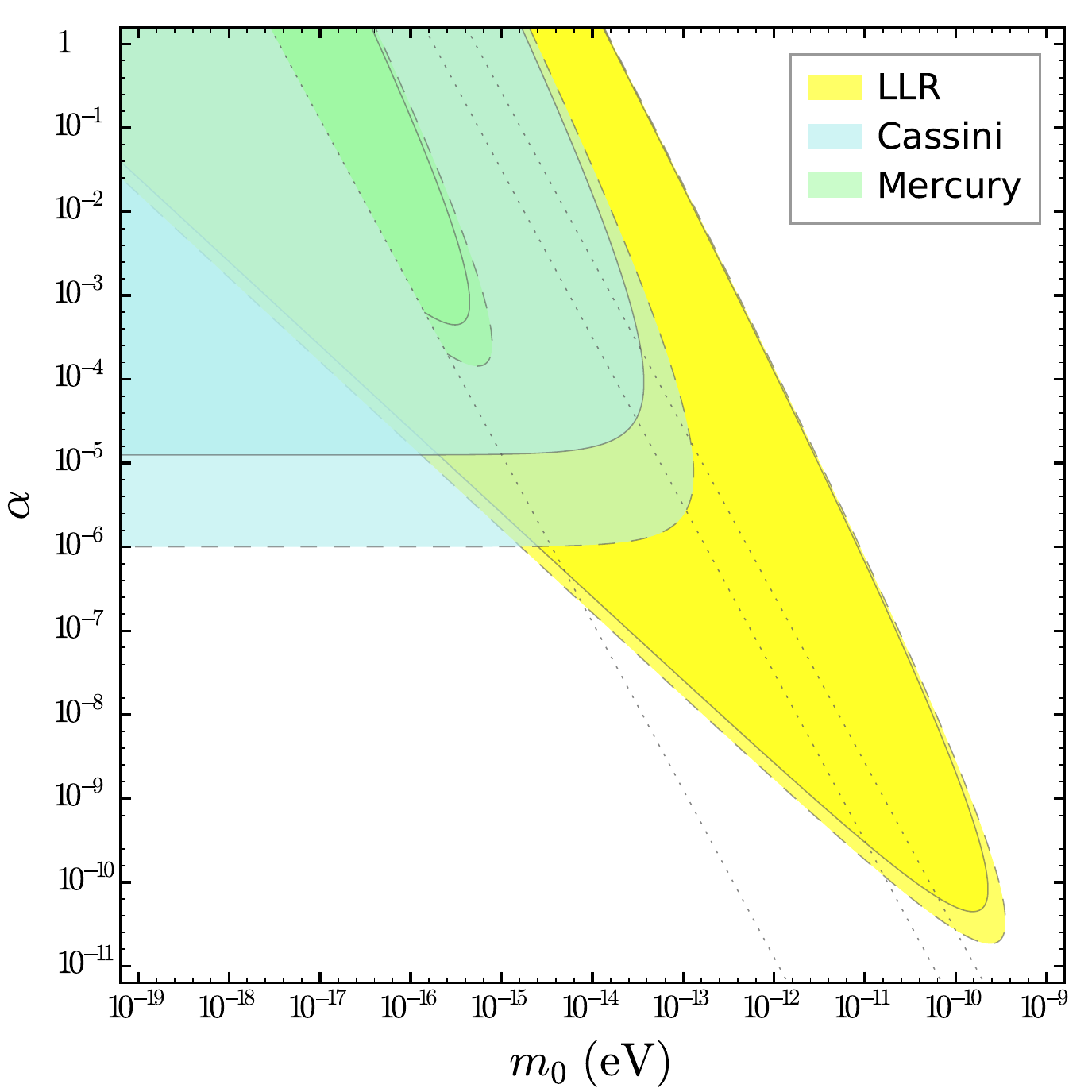}
\caption{
Solar system constraints on the coupling strength $\alpha$ of the fifth force as a function of its mediator mass $m_0$. 
The shaded regions are excluded by various experiments.
The dashed and solid lines represent $1\sigma$ CL and $2\sigma$ CL, respectively.
}
\label{fig1_alpha_m}
\end{figure}

\begin{figure}[!htbp]
\centering
\includegraphics[width=1\columnwidth]{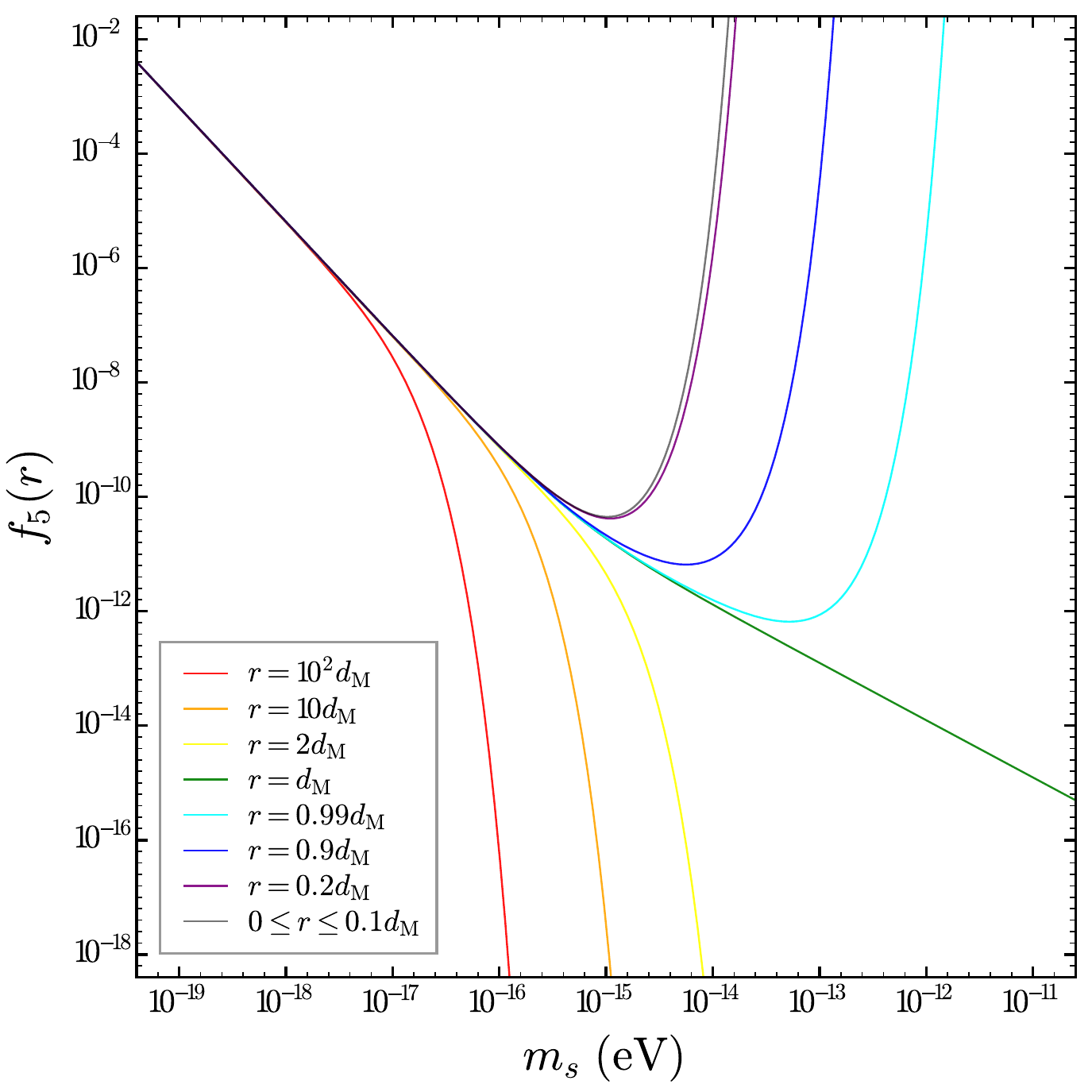}
\caption{$2\sigma$ CL upper bounds from the LLR observation on the strength ratio $f_5(r)$ at different distances as a function of effective mass $m_s$.}
\label{fig2_f5r_m}
\end{figure}

From Table \ref{tab_1}, we obtain the upper bounds
\begin{eqnarray}\label{precession_llr}
\frac{\delta\omega}{\omega} \le 0.5\times10^{-11}  
\quad{\rm and}\quad
\frac{\delta\omega}{\omega} \le 1.21\times10^{-11} 
\end{eqnarray}
at $1\sigma$ and $2\sigma$ confidence level (CL), the lower bounds
\begin{eqnarray}\label{gamma_lb}
\gamma \ge 0.999998  
\quad{\rm and}\quad
\gamma \ge 0.999975 
\end{eqnarray}
at $1\sigma$ and $2\sigma$ CL, and the upper bounds
\begin{eqnarray}\label{beta_ub}
\beta \le 1.000037 
\quad{\rm and}\quad
\beta \le 1.000115 
\end{eqnarray}
at $1\sigma$ and $2\sigma$ CL.
Using these above bounds and Eqs.~\eqref{f5_precession}, \eqref{gamma_a} and \eqref{beta_a}, we derive the observationally allowed range of the coupling strength $\alpha$ for each value of the mediator mass $m_0$, shown in Fig.~\ref{fig1_alpha_m}.
Using Eqs.~\eqref{f5_r} and \eqref{f5_precession}, the upper bounds on $\delta\omega/\omega$ in Eq.~\eqref{precession_llr} can be translated into the upper bounds on $f_5(r)$ for each value of the effective mass $m_s$, shown in Fig.~\ref{fig2_f5r_m}.
Note that, here, the upper bound of $\gamma$ and the lower bounds of $\delta\omega/\omega$ and $\beta$ cannot be used to constrain the theory, because of the conclusions \ref{item_2} and \ref{item_3}. 

\begin{table}[!htbp]
\centering
\caption{$2\sigma$ CL bounds on the coupling strength $\alpha$ of the fifth force for different values of its mediator mass $m_0$.}
\label{tab_2}
\resizebox{0.48\textwidth}{1.35cm}{
\begin{tabular}{p{1.8cm} < {\raggedright} p{1.65cm}< {\raggedright} p{2.3cm} <{\raggedright} p{2.15cm} < {\raggedleft}}
\hline\hline
$m_0 (\rm eV)$   &  $m_0^{-1} (\rm m)$ &  \qquad~~$\alpha$     &     Experiments  \\ 
\hline
$=0$                 &  $\infty$                &  $[0, 1.25\!\times\!10^{-5}]$  &    Cassini   \\
$\le10^{-15.7}$   &  $\ge10^{9.00}$      &  $[0, 1.25\!\times\!10^{-5}]$  &  Cassini \& LLR \\
$\ge10^{-9.65}$ &  $\le10^{2.95}$       &  $[0, +\infty)$              &    LLR  \\
$\ge10^{-13.4}$  &  $\le10^{6.70}$       &  $[0, +\infty)$              &    Cassini  \\
$\ge10^{-15.4}$  &  $\le10^{8.70}$       &  $[0, +\infty)$              &    Mercury  \\
\hline
\end{tabular}
}
\resizebox{0.48\textwidth}{0.35cm}{
\begin{tabular}{p{2.6cm} < {\raggedright} p{3.7cm} <{\raggedright} p{1.85cm} < {\raggedleft}}
$0\le\alpha\le10^{-10.4}$   &  \qquad~  $m_0(\rm eV) \in [0, +\infty)$       &   LLR \\ 
\hline\hline
\end{tabular}
}
\end{table}

In Fig.~\ref{fig1_alpha_m}, the yellow, cyan and green regions are excluded by the LLR measurement, the Cassini experiment, and the Mercury anomalous precession, respectively.
The three parallel dotted lines from right to left indicate the curves $m_s d_i = 1$ where $d_i =$ 0.00256AU, 0.00744AU, and 0.3707AU (see Table \ref{tab_1}) are the interaction distances of the LLR, Cassini, and Mercury experiments, respectively.
In the yellow regions, the dashed and solid lines are obtained by equating the $1\sigma$ and $2\sigma$ limits of Eqs.~\eqref{precession_llr} with Eq.~\eqref{f5_precession}, respectively, and plotting the corresponding contour in the parameter space $(\alpha, m_0)$.
The dashed and solid lines of the cyan regions are obtained in a similar way using Eqs.~\eqref{gamma_lb} and \eqref{gamma_a}.
In the green regions, the dashed and solid lines are obtained in a similar way using Eqs.~\eqref{beta_ub} and \eqref{beta_a}, and the upper right regions of the dotted line indicate $m_s d_i > 1$ ($d_i=0.3707 \rm AU$), in which Eq.~\eqref{beta_a} holds.
Clearly, the Mercury constraint on the parameter space $(\alpha, m_0)$ is the weakest one of all.

Table~\ref{tab_2} records several special points in Fig.~\ref{fig1_alpha_m}.
From Table~\ref{tab_2} we obtain the upper bound of $\alpha \le 1.25 \times 10^{-5}$ at $2\sigma$ CL for $m_0=0$.
This upper bound on $\alpha$ can be translated into the familiar bound $\omega \ge 40000$ in massless Brans-Dicke theory \cite{Brans:1961aa,Will:2014aa} by setting $F(\phi)=\phi$ and $W(\phi)=\omega(\phi)/\phi$ and substituting them into Eq.~\eqref{alpha_cp}.
We can see that at $2\sigma$ CL all values of $\alpha$ are allowed for $m_0 \ge 10^{-9.65} \rm eV$ in the LLR constraint, for $m_0 \ge 10^{-13.4} \rm eV$ in the Cassini constraint, and for $m_0 \ge 10^{-15.4} \rm eV$ in the Mercury constraint.
At $2\sigma$ CL, the horizontal coordinate of the intersection between $\delta\omega_5/\omega$ and $\gamma$ in Eqs.~\eqref{f5_precession} and \eqref{gamma_a} is $m_0 = 10^{-15.70} \rm eV$.
We can see that in the current precision the range of values of $\alpha$ for $m_0 \le 10^{-15.70} \rm eV$ is exactly the same as that of $m_0=0$.
In other words, the upper bound on $\alpha$ is no longer sensitive to values of $m_0$, when $m_0 \le 10^{-15.70} \rm eV$ or $m_0^{-1}\ge 10^{9} \rm m$, which is comparable to the interaction distance of $0.00744 \rm AU$ in Table~\ref{tab_1}.

Fig.~\ref{fig2_f5r_m} shows the constraints on the strength ratio $f_5(r)$ of the fifth force to gravitational force from the LLR observation searching for the anomalous precession of the Moon.
The curves represent the $2\sigma$ CL upper bounds on the strength ratio $f_5(r)$ as a function of effective mass $m_s$ at different distances $r/d_{\rm M} = $ 100(red), 10(orange), 2 (yellow), 1 (green), 0.99 (cyan), 0.9 (blue), 0.2 (purple), and $\in[0, 0.1]$ (grey).
Here, the effective mass $m_s$ is defined in Eq.~\eqref{ms_m0}, and $d_{\rm M}=0.00256 \rm AU$ is the semi-latus rectum of the Moon's orbit (see Table \ref{tab_1}).
We find that the $2\sigma$ CL upper bounds of $f_5(r\le0.1d_{\rm M})$ are approximately the same as that of $f_5(0)$ (see the grey curve).
As the distance $r$ decreases, the excluded parameter space decreases, and then the constraints on the strength ratio $f_5(r)$ become weak.
Therefore, the excluded parameter space of $r=0$ is located entirely in the excluded parameter space of $r>0$.
The LLR observation places much tighter fifth force constraint on large scales than on small scales.

\section{Conclusions}\label{sec8_concl}
General free scalar-tensor gravity with three free coupling functions is a natural extension to GR.
One of the most important consequences of this extension to GR is the emergence of a Yukawa-type fifth force.

In this paper, we explored the impact of the Yukawa-type fifth force in the theory on the PN metric and the orbital precession and derived in detail the PPN parameters $\gamma$ and $\beta$ and the fifth force-induced perihelion precession rate $\delta\omega/\omega$, respectively.
These parameters depend not only on the interaction distance from the test mass to the gravitational source, but also on the coupling strength and mediator mass of the Yukawa-type fifth force.
We found that if the fifth force vanishes then all these parameters reduce to their values in GR.
Our further analysis showed that the Yukawa-type fifth force in the theory is surprisingly always attractive, which leads to $\delta\omega/\omega>0$, $\gamma < 1$, and at large distances $\beta>1$.
In addition, this class of theories cannot completely substitute the role of dark matter in describing galactic dynamics.

Finally, we used solar system experiments (LLR, Cassini, and Mercury precession) to place stringent constraints on the parameter space of the attractive fifth force, and derived the upper bounds on the fifth force to gravitational force (scalar to tensor force) strength ratio at different scales with the LLR observation.
It turned out that the Mercury constraint is the weakest one of all, for small mediator masses the LLR constraint is not competitive with the Cassini constraint, but as the mediator mass increases the LLR constraint becomes tighter and tighter until it reaches the tightest constraint. 
Furthermore, we found that the upper bounds on the scalar to tensor force strength ratio become tighter as its interaction distance increases.

In this paper, all coupling functions are free, but the real physical world is usually governed by symmetry constraints.
So, in future work, we will study the general unfree scalar-tensor gravity constrained by theoretical considerations, which gives rise to extremely interesting results.

\section*{Acknowledgments}
This work is supported by the National Natural Science Foundation of China (NSFC) Grant No.11903033, No.12247103, No.12003029, No.12261131497, the National Key R\&D Program of China (2021YFC2203100), and the Fundamental Research Funds for the Central Universities.

\appendix
\section{PPN formalism}\label{app1_ppn}
The PPN formalism was inspired by the earlier work of Eddington \cite{Eddington:1923ty}, Robertson \cite{Robertson:1962ub} and Schiff \cite{Schiff:1967uk}.
Nordtvedt \cite{Nordtvedt:1968ab} and Will \cite{Will:1971aa} generalized the framework to a system of gravitating point masses and a perfect fluid, respectively.

In the PPN formalism, the metric for a perfect fluid is parametrized as follows \cite{Will:1993aa,Will:2014aa}:
\begin{eqnarray}\label{PPN_metric}
g_{00} &=& - 1 + 2 U - 2 \beta U^2 - 2 \xi \Phi_W      \nonumber\\
          &&   + (2 \gamma + 2 + \alpha_3 + \zeta_1 - 2 \xi ) \Phi_1     \nonumber\\
          &&   + 2 (3 \gamma - 2 \beta + 1 + \zeta_2 + \xi) \Phi_2        \nonumber\\
          &&   + 2 (1 + \zeta_3) \Phi_3 + 2 (3 \gamma + 3 \zeta_4 - 2 \xi)
                     \Phi_4 - (\zeta_1 - 2 \xi ) {\cal A}        \nonumber\\
          &&   - (\alpha_1 - \alpha_2 - \alpha_3) w^2 U -
                     \alpha_2 w^i w^j U_{ij}       \nonumber\\
          &&   + (2 \alpha_3 - \alpha_1 ) w^i V_i + {\cal O} (v^6),     \nonumber\\
g_{0i} &=& - \frac{1}{2} (4 \gamma + 3 + \alpha_1 -
                   \alpha_2 + \zeta_1 - 2 \xi ) V_i        \nonumber\\
         &&   - \frac{1}{2}(1 + \alpha_2 - \zeta_1 + 2 \xi) W_i -
                    \frac{1}{2} ( \alpha_1 - 2 \alpha_2 ) w^i U         \nonumber\\
         &&   - \alpha_2 w^j U_{ij} + {\cal O} (v^{5}),          \nonumber\\
g_{ij}  &=&    (1 + 2 \gamma U) \delta_{ij} + {\cal O} (v^4) \;,
\end{eqnarray}
where $w^i$ is coordinate velocity of the PPN coordinate system relative to the mean rest-frame of the universe.
The coefficients $\gamma$, $\beta$, $\xi$, $\alpha_1$, $\alpha_2$, $\alpha_3$, $\zeta_1$, $\zeta_2$, $\zeta_3$ and $\zeta_4$ are the PPN parameters, whose significances are as follows: $\gamma$ measures the amount of space-curvature produced by unit rest mass, $\beta$ measures the non-linearity in the superposition law for gravity, $\xi$ measures the violation of local position invariance, $\alpha_1$, $\alpha_2$ and $\alpha_3$ measure the violation of local Lorentz invariance, and $\alpha_3$, $\zeta_1$, $\zeta_2$, $\zeta_3$ and $\zeta_4$ measure the violation of total energy-momentum conservation.

The metric potentials $U$, $U_{ij}$, $\Phi_W$, $\cal A$, $\Phi_1$, $\Phi_2$, $\Phi_3$, $\Phi_4$, $V_i$ and $W_i$ are constructed as follows \cite{Will:1993aa,Will:2014aa}:
\begin{eqnarray}\label{Metric_potentials}
U         &=&  \int \frac{\rho'}{|{\bf x}-{\bf x}'|} \, d^3x',    \nonumber\\ 
U_{ij}    &=&  \int \frac{\rho' (x-x')_i (x-x')_j} {|{\bf x}-{\bf x}'|^3} \, d^3x',    \nonumber\\
\Phi_W &=&  \int \frac{\rho' \! \rho'' ({\bf x}-{\bf x}')} {|{\bf x}-{\bf x}'|^3} \! \cdot \!
                    \left(\! \frac{{\bf x}' \!-\! {\bf x}''}{|{\bf x}-{\bf x}''|} -
                     \frac{{\bf x}-{\bf x}''}{| {\bf x}'-{\bf x}''|} \!\right) \! d^3x'  d^3x'',   \nonumber\\
{\cal A} &=&  \int \frac{\rho' [{\bf v}' \cdot ({\bf x}-{\bf x}')]^2} {|{\bf x}-{\bf x}'|^3} \, d^3x',    \nonumber\\
\Phi_1   &=&  \int \frac{\rho' v'^2}{|{\bf x}-{\bf x}'|} \, d^3x',     \nonumber\\
\Phi_2  &=&  \int \frac{\rho' U'}{|{\bf x} - {\bf x}'|} \, d^3x',       \nonumber\\
\Phi_3  &=&  \int \frac{\rho' \Pi'}{|{\bf x}-{\bf x}'|} \, d^3x',       \nonumber\\
\Phi_4  &=&  \int \frac{p'}{|{\bf x}-{\bf x}'|} \, d^3x',                 \nonumber\\
V_i       &=&  \int \frac{\rho' v_i'}{|{\bf x}-{\bf x}'|} \, d^3x',        \nonumber\\
W_i      &=&  \int \frac{\rho' [{\bf v}' \cdot ({\bf x}-{\bf x}')](x-x')_i} {|{\bf x}-{\bf x}'|^3} \, d^3x' \;,
\end{eqnarray}
where $\rho$, $p$ and $\Pi$ are respectively the rest mass density, pressure and specific internal energy of the perfect fluid (see the energy-momentum tensor of the perfect fluid in Eq.~\eqref{Tuv_per_fluid}).
The metric potentials satisfy the following simple and useful relationships \cite{Will:1993aa,Will:2014aa}
\begin{eqnarray}\label{Potential_relationship}
\nabla^2 U = - 4\pi \rho ,                \label{Rel_1}\\
\nabla^2 \Phi_1 = - 4\pi \rho v^2 ,    \label{Rel_2}\\
\nabla^2 \Phi_2 = - 4\pi \rho U ,      \label{Rel_3}\\
\nabla^2 \Phi_3 = - 4\pi \rho \Pi ,    \label{Rel_4}\\
\nabla^2 \Phi_4 = - 4\pi p ,              \label{Rel_5}\\
\nabla^2 V_i = - 4\pi \rho v_i ,          \label{Rel_6}\\
\nabla^2 (W_i - V_i) = 2 U_{,0i},         \label{Rel_7}\\
U\nabla^2 U = \nabla^2 \Phi_2 ,        \label{Rel_8}\\
(\nabla U)^2 = \nabla^2 (\frac{1}{2}U^2 - \Phi_2) \;.    \label{Rel_9}    
\end{eqnarray}

\bibliography{gfst_f5}

\end{document}